\documentclass[tightenlines,draft,aps,twocolumn,showpacs,floatfix]{revtex4}
\usepackage{psfig}
\newcommand{\be}{\begin{equation}}
\newcommand{\ee}{\end{equation}}
\newcommand{\ba}{\begin{eqnarray}}
\newcommand{\ea}{\end{eqnarray}}

\begin{document}      
%

%

\preprint{\vbox{\baselineskip=14pt%
   \rightline{???-HEP-???}      
   \rightline{July 2002}      
}}      
      
\title{
\[ \vspace{-2cm} \]
\noindent\hfill\hbox{\rm  SLAC-PUB-9373} \vskip 1pt
\noindent\hfill\hbox{\rm hep-ph/0208115} \vskip 10pt
Pseudoscalar Higgs boson production at hadron colliders in NNLO QCD
}      
\author{Charalampos Anastasiou and Kirill Melnikov}      
\affiliation{Stanford Linear Accelerator Center, Stanford
  University,Stanford,
  CA 94309, U.S.A.}         

\begin{abstract}    
We compute the total cross-section for direct production of the 
pseudoscalar  Higgs boson in hadron collisions at 
next-to-next-to-leading order (NNLO) in perturbative QCD. 
The ${\cal O}(\alpha_s^2)$ QCD corrections increase the NLO production 
cross-section by approximately $20-30$ per cent.
\end{abstract}      
\maketitle     

\section{Introduction}
\label{sec:intro}    
Supersymmetry is one of the most popular extensions of the  
Standard Model (SM).  Generically, supersymmetric theories predict
a very rich spectrum  of elementary particles. In particular, the Higgs boson 
sector of the Minimal Supersymetric Standard Model (MSSM) consists of two 
complex Higgs  doublets.  After electroweak symmetry breaking, 
three Higgs fields are absorbed by the $W^{\pm}$ and $Z$ bosons into 
their longitudinal degrees of freedom; the remaining five degrees 
of freedom are physical Higgs bosons. 
In addition to the standard Higgs boson $(h)$, 
a heavier neutral Higgs boson $(H)$, two charged scalar Higgs bosons 
$(H^\pm)$,  and a neutral pseudoscalar Higgs boson $(A)$ appear in 
the spectrum.

The tree-level masses of the Higgs bosons in the MSSM are usually 
described in terms of two independent parameters: the mass of the 
pseudoscalar Higgs boson $m_A$  and the ratio of the vacuum expectation values of the 
two Higgs doublets $\tan \beta = v_1/v_2$. Currently, these parameters are 
restricted by the experiments at LEP which set a lower bound 
$m_A > 91.9~{\rm GeV}$ and exclude the values $0.5 < \tan \beta <
2.4$~\cite{lep,lepwg}. Future searches for the pseudoscalar Higgs boson 
will be carried out at the Tevatron and at the LHC. It is therefore
important to obtain a reliable theoretical estimate of its production 
cross-section in hadron collisions.  

The pseudoscalar Higgs boson does not couple to the gauge bosons 
at tree level;  the major  
mechanisms for producing it are the gluon-gluon fusion mechanism 
$gg \to A$ and the  associated  production process $gg,qq \to A qq$. The relative significance of 
the production channels depends on the mass of the Higgs boson and the 
value of  $\tan \beta$. For larger values of  $\tan \beta$, the coupling of 
the pseudoscalar Higgs boson to up quarks,  $g_{\rm up}  \sim m_u/\tan \beta$, 
decreases while the coupling of the Higgs boson to down quarks,  
$g_{\rm down} \sim m_d \tan \beta$, increases. As a consequence, 
the phenomenology of the axial Higgs for large values of $\tan \beta$ 
is very different from the phenomenology of the SM Higgs boson. 
For the $(t,b)$ family, the Higgs boson interaction with bottom quarks 
dominates for values of $\tan \beta \ge 10$.

Gluon-gluon fusion through a quark triangle-loop   
is the dominant production channel of the light pseudoscalar Higgs boson 
in hadron collisions; there is no squark-loop contribution to the $ggA$ 
coupling. In this Letter we study the gluon-gluon 
fusion cross-section for small and moderate values of $\tan \beta$.  
We can then neglect the contribution of bottom quarks and focus 
on the production of the pseudoscalar Higgs boson through the top-quark 
loop. In addition, we consider small values of the pseudoscalar Higgs 
boson mass, $m_A \le 300~{\rm GeV}$. Since $m_A \ll 2m_t$, where $m_t$ is the 
mass of the top-quark, the interaction of the pseudoscalar Higgs boson with 
gluons and light quarks can be described by the effective 
Lagrangian~\cite{Chetyrkin:1998mw}
\be
{\cal L} = \frac{A}{v \tan \beta }\left [ \tilde C_1~G_{\mu \nu}\tilde G^{\mu \nu} 
+ \tilde C_2 \partial_\mu J_5^{\mu}
\right],
\label{eq1}
\ee
where $G_{\mu \nu}$ is the color field-strength tensor and 
$$
\tilde G_{\mu \nu} = 
\epsilon_{\mu\nu\alpha\beta}~G^{\alpha \beta},~~~
J_5^{\mu} = \sum \limits_{i}^{n_f} \bar q_i \gamma_\mu \gamma_5 q_i.
$$
The Wilson coefficients $\tilde C_1$ and $\tilde C_2$ are given by
\ba
&& \tilde C_1 = -\frac{\alpha_s(\mu)}{16\pi}, \nonumber \\
&& \tilde C_2 = \left ( \frac{\alpha_s(\mu)}{\pi} \right )^2 
\left ( \frac{1}{8} - \frac{1}{4} L_t \right),
\ea
where $\alpha_s(\mu)$ is the $\overline {\rm MS}$ coupling constant, 
$n_f$ is the number of massless quark flavors and $L_t = \log(\mu^2/m_t^2)$.

The renormalization in higher orders of the perturbation theory 
of the effective Lagrangian in Eq.~(\ref{eq1}) is subtle.
Because of the axial anomaly, the derivative of the axial current of 
light quarks, $\partial_\mu J_5^{\mu}$, mixes under renormalization 
with the operator $G_{\mu \nu}~\tilde G^{\mu \nu}$. 
The renormalization procedure must preserve the anomaly relation
\be
\partial_\mu J_5^{\mu} = 
\frac{\alpha_s(\mu)n_f}{8\pi}~G_{\mu \nu}\tilde G^{\mu \nu}. 
\ee 
A  detailed discussion of the effective 
Lagrangian and its renormalization can be found in 
Ref.~\cite{Chetyrkin:1998mw}. 

The Levi-Civita tensor $\epsilon_{\mu\nu \alpha \beta}$ and the $\gamma_5$
Dirac matrix are four-dimensional and their treatment in dimensional 
regularization is a delicate problem. 
For calculational convenience we use the approach suggested 
in Ref.~\cite{Larin:tq}. According to this prescription 
the $\gamma_5$ matrix is represented as 
$\gamma_5=i/24~\epsilon_{\mu \nu \alpha \beta} \gamma^{\mu} \gamma^{\nu} \gamma^\alpha \gamma^\beta$ and the axial-vector current as 
$J_5^{\mu} = 1/2 \bar \psi (\gamma _\mu \gamma_5 - \gamma_5 \gamma_\mu)\psi$.
With this substitution we can factor out the product of two Levi-Civita  
tensors from the production cross-section and evaluate it in terms 
of $d=4-2\epsilon$ dimensional metric tensors $\delta_{\mu \nu}$, using
\be
\epsilon_{\mu_1 \mu_2 \mu_3 \mu_4} \epsilon^{\nu_1 \nu_2 \nu_3 \nu_4} 
= -\left |
\begin{array}{cccc}
\delta_{\mu_1}^{\nu_1} & \delta_{\mu_1}^{\nu_2} & \delta_{\mu_1}^{\nu_3}& \delta_{\mu_1}^{\nu_4}\\
\delta_{\mu_2}^{\nu_1} & \delta_{\mu_2}^{\nu_2} & \delta_{\mu_2}^{\nu_3}& \delta_{\mu_2}^{\nu_4}\\
\delta_{\mu_3}^{\nu_1} & \delta_{\mu_3}^{\nu_2} & \delta_{\mu_3}^{\nu_3}& \delta_{\mu_3}^{\nu_4}\\
\delta_{\mu_4}^{\nu_1} & \delta_{\mu_4}^{\nu_2} & \delta_{\mu_4}^{\nu_3}& \delta_{\mu_4}^{\nu_4}\\
\end{array}
\right |.
\ee

We have  checked that the above prescription is consistent with 
the renormalization procedure of Ref.~\cite{Chetyrkin:1998mw} by calculating 
the decay rate of the pseudoscalar Higgs boson through NNLO. Our 
results are in agreement with the expressions for the decay rate 
given in Ref.~\cite{Chetyrkin:1998mw}, where a four-dimensional treatment of 
the Levi-Civita tensors was employed. 

In this Letter we present the NNLO QCD corrections to the pseudoscalar Higgs
boson production cross-section in hadron collisions. 
Various partonic processes contribute to the cross-section at this 
order. Specifically, we have to compute:
a) virtual corrections to $gg \to A$, $q\bar q \to A$ up to 
${\cal O}(\alpha_s^2)$;
b) virtual corrections to single real emission processes $gg \to Ag$, 
$qg \to Aq$, $\bar q g \to A \bar q $, $q\bar q \to A g$, 
up to ${\cal O}(\alpha_s)$;  and c) double real emission processes 
$gg \to Agg$, $gg \to Aq \bar q$, $qg \to Aqg$, $q \bar q \to Agg$, 
$q \bar q \to A q \bar q$. 
We evaluate the above corrections using the method introduced in 
Ref.~\cite{babis} for the algorithmic evaluation of phase-space integrals. 

\section{Partonic Cross-Sections}
\label{sec:partonic}
In this section we present analytic expressions for the partonic 
cross-sections $i +j \to A +X$, where $i,j=q, \bar q, g$. We write 
\be
\hat \sigma_{ij} = \sigma_0^{(A)} \left [ \phi_{ij}^{(0)} +
\left ( \frac{\alpha_s}{\pi} \right ) \phi_{ij}^{(1)} 
+ \left ( \frac{\alpha_s}{\pi} \right )^2 \phi_{ij}^{(2)} 
\right ],
\ee
where 
\be
\sigma_0^{(A)} = 
\frac{\pi}{256 v^2 \tan^2 \beta} \left (\frac{\alpha_s}{\pi} \right )^2. 
\ee
The coefficients $\phi^{(k)}_{ij}$ are very similar to the coefficients 
$\eta^{(k)}_{ij}$ of the perturbative expansion of the partonic 
cross-sections for the production of the scalar Higgs boson~\cite{babis}.  
We can then write
\be
\phi^{(k)}_{ij} \equiv \delta \phi^{(k)}_{ij} + \eta^{(k)}_{ij}.
\ee
For convenience, we present here the difference 
$\delta \phi^{(k)}_{ij} = \phi^{(k)}_{ij} - \eta^{(k)}_{ij}$. The 
$\eta^{(k)}_{ij}$ terms are listed in Section~IV of Ref.~\cite{babis}.

We set the renormalization and the factorization scales equal to the mass 
of the Higgs boson $m_A$.
At LO we find
\begin{equation}
\delta \phi^{(0)}_{ij} = 0.
\end{equation}
At NLO we obtain
\begin{equation} 
\delta \phi ^{(1)}_{gg} 
= \frac{1}{2} \delta(1-x),
\end{equation}
and 
\begin{equation}
\delta \phi^{(1)}_{qg,q\bar q} = 0.
\end{equation}
The NLO terms are in agreement with the results of Ref.~\cite{spira}.

At NNLO the differences $\delta \phi^{(2)}_{ij}$ are very simple. For 
both  $\phi^{(2)}_{ij}$ and  $\eta^{(2)}_{ij}$ we find the same 
polylogarithmic terms; the differences contain simple 
logarithms and the Riemann $\zeta_2$ constant.  We obtain
\ba
 \delta \phi^{(2)}_{gg} &=&
 \left[ \left(  \frac{{\it L_t}}{3}-\frac{21}{16} \right)  \delta \left( 1-x
 \right) +\frac{2}{3} x \ln^2  \left( x \right)+x \ln 
 \left( x \right) \right. \nonumber \\
&& \left. 
-  \frac{1}{6} \left( 10 x-1 \right)  \left( x-1 \right) 
 \right] {\it n_f}+6 {\left[\frac{\ln(1-x)}{(1-x)} \right]_{+}} 
 \nonumber \\ &&
+ \left( 3 \zeta_2  +\frac{1939}{144}-\frac{19}{8} {\it L_t} \right) 
\delta \left( 1-x \right)  \nonumber \\
&& -6 x \left( {x}^{2}-x+2 \right) \ln  \left( 1-
x \right) -9 x \ln^2 \left( x \right) \nonumber \\
&& +\frac{3}{2} 
\left( 2 {x}^{4}-4 {x}^{3}+13 {x}^{2}+x-10 \right) {\frac {  \ln 
\left( x \right) }{(x-1)}}
\nonumber \\
&& +\frac{1}{4}  \left( x-1 \right)  \left( 11 {x}^{2}
+35 x-154 \right),
\ea
\ba 
 \delta \phi^{(2)}_{qg} &=&
\frac{2}{3} \left( -2 x+2+ {x}^{2} \right) \ln  \left( 1-x \right) -{
\frac {28}{9}} x \ln^2  \left( x \right) 
\nonumber \\
&& + \left[ {
\frac {22}{3}}+10 x-\frac{ {x}^{2}}{3} \right] \ln  \left( x \right) +{
\frac {17}{6}} {x}^{2}-{\frac {191}{9}} x
\nonumber \\
&& +{\frac {337}{18}},
\ea
\ba
 \delta \phi^{(2)}_{q\bar q} &=&
\left[ -{\frac {32}{27}}  x \ln  \left( x \right) +{\frac {16}{27}} 
 \left( x^2-1 \right)  \right] {\it n_f}
 \nonumber \\
&& +
{\frac {32}{27}}  x \ln^2  \left( x \right) +{\frac {
32}{27}} { \left( 3+8 x \right) \ln  \left( x \right) }
\nonumber \\
&& -
\frac{16}{27} \left( x-1 \right)  \left( {x}^{2}+10 x+11
 \right),  
\ea
\ba
 \delta \phi^{(2)}_{qq'} &=&
-{\frac {16}{9}} x  \ln^2  \left( x \right) +
 \left[ \frac{16}{3} x+{\frac {32}{9}} \right] \ln  \left( x \right) 
 \nonumber \\
&& +{\frac 
{8}{9}}  \left( x-1 \right)  \left( x-11 \right),  
\ea
and
\ba 
 \delta \phi^{(2)}_{qq} &=&
-{\frac {64}{27}} x \ln^2  \left( x \right) +
 \left[ {\frac {32}{9}}+{\frac {176}{27}} x \right] \ln  \left( x
 \right) 
 \nonumber \\
&& +{\frac {8}{27}}  \left( x-1 \right)  \left( 3 x-37
 \right). 
\ea
The above results are valid if the renormalization and factorization 
scales are equal to the mass of the Higgs boson, $\mu_f = \mu_r = m_A$. 
The complete functional dependence of the partonic cross-sections 
on these scales can be easily restored by solving the DGLAP equation 
and the renormalization group equation and using the above expressions as 
the boundary conditions. This procedure is outlined in Ref.~\cite{babis}.

\section{Numerical Results}
\label{sec:numerics}

\begin{figure}[h]
\begin{picture}(100,0)
\put (-30,0) {$\tan^2\beta~\sigma(p p \to A + X)~[{\rm pb}],~~\sqrt{s} = 14~{\rm TeV}$}
\put (20,-170) {$m_A,~{\rm GeV}$}
\end{picture}
\psfig{figure=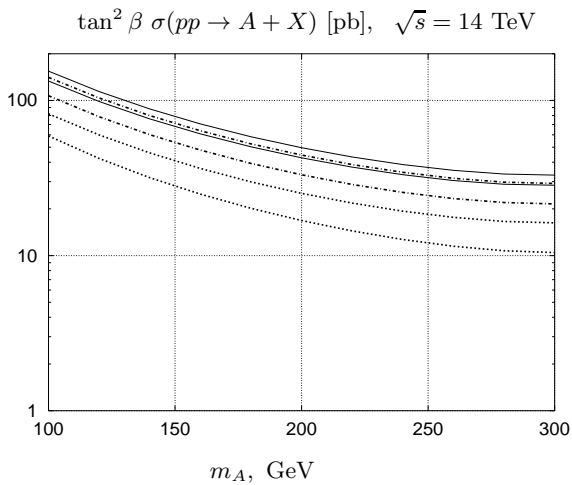,width=80mm}
\vspace*{0.5cm}
\caption{The pseudoscalar Higgs boson production cross-section at the LHC at 
leading (dotted), next-to-leading (dashed-dotted) and
 next-to-next-to-leading (solid)
order. The two curves for each case correspond to $\mu_r = \mu_f = m_A/2$ (upper) and $\mu_r = \mu_f = 2 m_A$ (lower).}
\label{fig:LHC}
\end{figure}

We now discuss the numerical impact of 
the NNLO corrections on the pseudoscalar Higgs boson production cross-section 
at the LHC and the Tevatron. To calculate the cross-section we must convolute 
the hard scattering partonic cross-sections of Section~\ref{sec:partonic} 
with the appropriate parton distribution functions, according to the 
factorization formula
\be
\sigma_{A} = x 
\sum \limits_{ij}^{} \left [
\bar f_i^{(h_1)} \otimes
\bar f_j^{(h_2)} \otimes ( \sigma_{ij}(z)/z ) \right ](x). 
\ee
Here $\bar f_i^{(h)}$ is the $\overline {\rm MS}$ distribution 
function of the parton $i$ in the hadron $h$,  
$\otimes$ denotes  the standard convolution,
\be
(f \otimes g )(x) \equiv \int_0^1 dy dz f(y) g(z) \delta(x-yz),
\ee
and $x = m_A^2/s$, where $s$ is the square of the total center of mass energy 
of the hadron-hadron collision. 

The complete NNLO parton distribution functions are not  yet
available. In Ref.~\cite{msrt} an approximate NNLO 
evolution~\cite{neervenvogt} has been implemented in order to 
determine the NNLO MRST parton distribution functions. We  use these 
approximate solutions 
for the numerical evaluation of the 
pseudoscalar Higgs boson production cross-section keeping the same 
initialization parameters as in Ref.~\cite{babis}.     

To demonstrate the convergence properties of the perturbative 
series for the  hadronic cross-section, we present the LO, NLO and 
NNLO results for both the LHC and the Tevatron. In order to improve upon 
the heavy-top quark approximation, we normalize our results to the 
leading-order cross-section with the exact dependence on $m_t$.   

The total cross-section for the LHC is shown in Fig.~\ref{fig:LHC}.
\begin{figure}[h]
\begin{picture}(100,0)
\put (-30,0) {$\tan^2\beta~\sigma(p p \to A + X)~[{\rm pb}],~~\sqrt{s} = 2~{\rm TeV}$}
\put (20,-170) {$m_A,~{\rm GeV}$}
\end{picture}
\psfig{figure=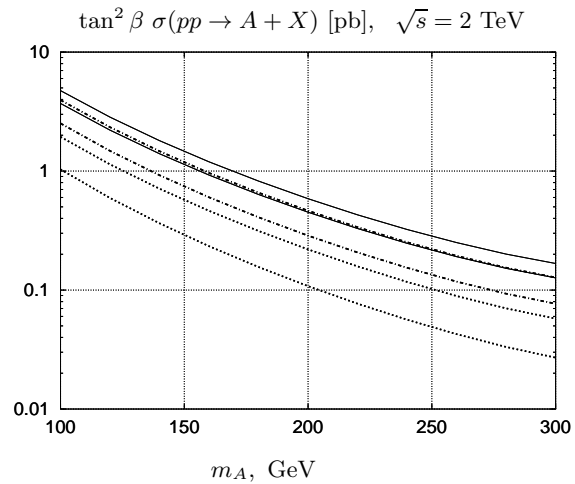,width=80mm}
\vspace*{0.5cm}
\caption{The pseudoscalar Higgs boson production cross-section at the 
Tevatron at leading (solid), next-to-leading (dashed) and 
next-to-next-to-leading (dotted) order. The two curves for each case 
correspond to $\mu_r = \mu_f = m_A/2$ (upper) and $\mu_r = \mu_f = 2 m_A$ (lower).}
\label{fig:Tevatron}
\end{figure}
From Fig.\ref{fig:LHC} we observe that the scale dependence 
of the Higgs production cross-section at NNLO is approximately 
$15\%$; this is a factor of two smaller than the NLO scale dependence 
and a factor of four less than the scale variation at LO. Despite the scale
stabilization, the corrections are rather large;
the NLO corrections increase the LO cross-section by about $70\%$, 
and the NNLO corrections further increase it  by approximately 
$30\%$. The $K$ factor, defined as the ratio of the NNLO cross-section 
and the LO cross-section, is  approximately two.
In Fig.\ref{fig:Tevatron} we plot the values of the Higgs production 
cross-section at the Tevatron. The NNLO $K$ factor is approximately 
three, and  the residual scale dependence is approximately $30\%$. 
The $K$ factors for the production of the pseudoscalar Higgs boson and the 
scalar Higgs boson~\cite{hkhard,babis} are comparable in magnitude; this is a 
consequence of the similarity of the corresponding partonic cross-sections 
as discussed in Section~\ref{sec:partonic}.

\begin{figure}[htb]
\begin{picture}(100,0)
\put (-30,0) {$\tan^2\beta~\sigma(p \bar p \to A + X)~[{\rm pb}],~~\sqrt{s} 
 = 14~{\rm TeV}$}
\put (30,-170) {$\mu,~{\rm GeV}$}
\end{picture}
\psfig{figure=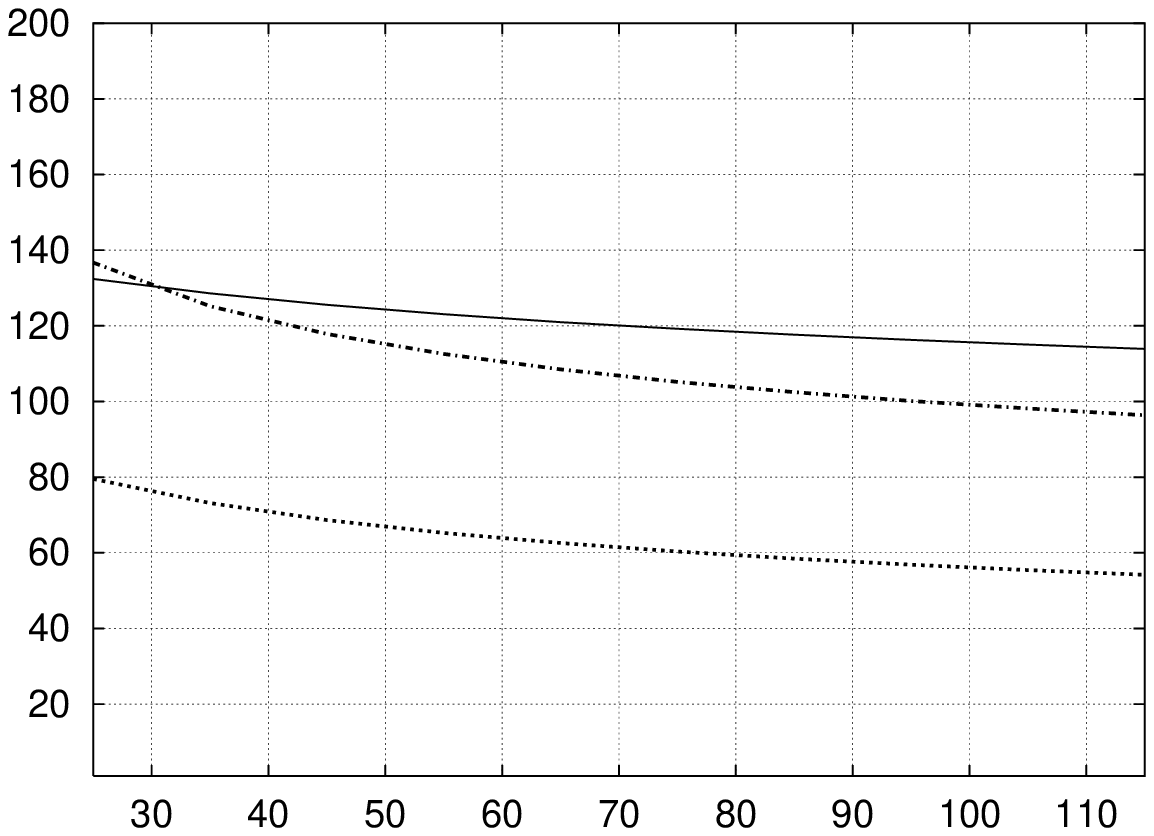,width=80mm}
\vspace*{0.5cm}
\caption{
The pseudoscalar Higgs boson production cross-section at the LHC at 
leading (dotted), next-to-leading (dashed) and next-to-next-to-leading (solid)
order as the function of factorization and renormalization scale $\mu$.
The mass of the Higgs boson is $115~{\rm GeV}$. }
\label{fig:plot2}
\end{figure}

In Ref.~\cite{babis} we have argued that since the dominant contribution 
to the integrated cross-section for the scalar Higgs boson comes from the 
region close to the Higgs boson production threshold, we should choose 
values of the scale $\mu$ which are smaller than the mass of the Higgs boson.  
This choice {\it decreases} the NNLO corrections and  the Higgs boson 
production cross-section {\it increases} as compared 
to conventional choice of the scales, $\mu_r = \mu_f = m_H$.
The production cross-section for the pseudoscalar Higgs boson exhibits the 
same behavior.  In Fig.~\ref{fig:plot2} we show an example,
where we plot the production cross-section for  $m_A = 115~{\rm GeV}$. 
We equate the renormalization and factorization 
scales and vary the factorization scale from $\mu = 25~{\rm GeV}$
up to the mass of the pseudoscalar Higgs boson. The plot illustrates that 
for smaller values of $\mu$, the NLO cross-section increases more 
rapidly than the NNLO cross-section, and the 
difference between the NLO and the NNLO results becomes smaller.
Therefore, the convergence of the perturbative series is improved 
for smaller values of the factorization scale. 
\\
\section{Conclusions}
\label{sec:conclusions}
In this Letter we presented the NNLO QCD corrections for the 
production cross-section of the pseudoscalar Higgs boson in hadron collisions.
Our results are valid in the heavy top-quark limit and for small to moderate 
values of $\tan \beta$.  

The analytic expressions which we presented here and those for 
the scalar Higgs boson~\cite{babis} are very similar.
In both cases, the QCD corrections are large and important for quantitative 
estimates of the hadronic cross-sections at the Tevatron and the LHC. 
The size of the NNLO corrections for the pseudoscalar Higgs boson indicates 
that the perturbative expansion of the production cross-section converges, 
albeit slowly. A similar convergence behavior was observed for the SM Higgs 
boson hadronic production cross-section~\cite{hkhard,babis}.

In order to verify the compatibility of the prescription of 
Ref.~\cite{Larin:tq} for the Levi-Civita tensor 
with the Wilson coefficients for the effective Lagrangian in Eq.~(\ref{eq1}),
derived in Ref.~\cite{Chetyrkin:1998mw}, 
we computed the pseudoscalar Higgs boson decay rate through NNLO in QCD. 
Our results are in complete agreement with the expressions for the decay rate 
in~\cite{Chetyrkin:1998mw}.

We note that the calculation reported in this Letter has been 
performed using the method of Ref.~\cite{babis} which combines the optical 
theorem with integration-by-parts reduction algorithms to achieve a
systematic evaluation of phase-space integrals.  As our calculation 
demonstrates, the method is general  and process-independent.  
We are confident that the same method will be very useful in 
studying  other processes of phenomenological  interest.

Finally, as we completed this manuscript, we became aware of a similar 
calculation by R. Harlander and W. Kilgore \cite{axialhk}.  
We have compared results for the partonic cross-sections 
and find complete agreement.

{\bf Acknowledgments:} 
We are grateful to Maria Elena Tejeda-Yeomans for  collaboration 
during the early stages of this work.  We thank Frank  Petriello for 
useful discussions. This research was supported by the DOE under grant
number DE-AC03-76SF00515.

\end{document}